\documentclass{article}
\usepackage{epsfig,amssymb,euscript}
\usepackage{amsmath}

\newcommand{\halpha}{\hat{\alpha}}

\def\be#1\ee{\begin{equation}#1\end{equation}}
\newcommand{\bea}{\begin{eqnarray}}
\newcommand{\eea}{\end{eqnarray}}
\newcommand{\ba}{\begin{array}}
\newcommand{\ea}{\end{array}}

\def\bbox{{\,\lower0.9pt\vbox{\hrule \hbox{\vrule height 0.2 cm
\hskip 0.2 cm \vrule height 0.2 cm}\hrule}\,}}
\newcommand{\dsl}{\pa \kern-0.5em /}

\def\ds{\raise.15ex\hbox{/}\kern-.57em\partial}
\def\Ds{\,\raise.15ex\hbox{/}\mkern-13.5mu D}

\begin{document}

\makeatletter
\renewcommand{\theequation}{\thesection.\arabic{equation}}
\@addtoreset{equation}{section} \makeatother

\baselineskip 18pt


\begin{titlepage}

\vfill

\begin{center}
\baselineskip=16pt
{\Large\bf Spontaneous Collapse Models on a Lattice}
\vskip 10.mm Fay Dowker$^{1}$ and Joe Henson$^{2}$\\
\end{center}  
\vfill
\par
\begin{center}
{\bf Abstract}
\end{center}
\begin{quote}
We present 
spontaneous collapse models of  field theories on a $1+1$ null lattice, 
in  which the causal structure of the lattice plays a central role.
Issues such as ``locality,'' ``non-locality'' and superluminal 
signaling are addressed in the context of the models which 
have the virtue of extreme simplicity. The formalism of the models 
is related to that of
the consistent histories approach to quantum mechanics. 


\end{quote}

\vfill 

\hrule width 5.cm \vskip 5mm {\small
\noindent $^1$ Blackett Laboratory, Imperial College, Prince Consort Road,
London SW7 2BZ, UK, and Perimeter Institute, 35 King Street North, Waterloo,
ON N2J 2W9, Canada. \\
\noindent $^2$ Department of 
Mathematics, University of California/San Diego, La Jolla,
CA 92093-0112, USA.\\ 
}
\end{titlepage}
\setcounter{equation}{0}

\section{Introduction}

The spontaneous collapse models of Ghirardi, Rimini and  Weber (GRW),
Pearle and others (see \cite{Bassi:2003gd} for a review)
represent a promising direction of research towards 
an observer independent theory of fundamental matter.
These models were first proposed in a non-relativistic 
framework and since then much attention has focused 
on the search for appropriately relativistic models.
This is not only important in its own right, but seems to be
a prerequisite to any hope of applying collapse model 
ideas to quantum gravity.
 
We present a simple collapse model for a field theory
defined on a $1+1$ null lattice. It is inspired by  
work by Samols \cite{Samols:1995zz} and by the 
interpretation of the GRW model \cite{Ghirardi:1986mt}, 
due to Bell \cite{Bell:1987}, in which 
it is the collapse centres that are the ``beables'' 
or ``real variables''.  In the cited work, 
Bell proved a result suggestive that Lorentz invariant collapse models 
can be formulated.  When the system of particles treated by the GRW 
model can be split into two non-interacting subsystems, the time 
evolution of one subsystem has no effect on the other, as in 
standard quantum mechanics. The work presented here can be considered as 
further support for Bell's expectation that a fully Lorentz invariant 
collapse model with the ontology he proposed can be 
constructed. In particular we prove, in the 
framework of the lattice collapse model, the analogue of 
Bell's result that, ``Events in one system, considered
separately, allow no inference [...] about external fields at work in the
other.''

Although the lattice collapse model is not itself Lorentz invariant, 
it is not unreasonable to hope that Lorentz invariance
will be attained in an appropriate
continuum limit.
 Moreover it is the view of some workers that the 
aspect of spacetime that is fundamental and survives its
encounter with its quantum nemesis is its causal structure
and, further, that this fundamental causal structure is discrete 
\cite{Bombelli:1987aa}. If one  is looking for a development 
of quantum theory suited to such beliefs about quantum gravity,
our model has many attractive features: it is discrete, there 
is a causal structure, and 
there is a local evolution rule tied to  that causal structure.

\section{1+1 lattice quantum field theory}

We briefly review the basics of light cone lattice field 
theory in 1+1 dimensions, introduced in the 
study of integrable models \cite{Destri:1987ze}.
We follow the presentation of Samols \cite{Samols:1995zz} 
of this ``bare bones'' local quantum field theory.
Spacetime is a $1+1$ null 
lattice, periodically identified in space of width 2N. 
We label the links of the lattice L or R depending on 
whether they are left or right moving null rays. A 
spacelike surface, $\sigma$, is given by the set of links cut by 
the surface and is specified completely by the position of 
an initial 
link and a sequence of $N$ $R$'s and $N$ $L$'s
labeling the links it cuts successively, moving from left to 
right, starting with the initial link. An example 
is shown in figure 1, taken from Samols' paper.   

\begin{figure}[thb]
\centerline{\epsfbox{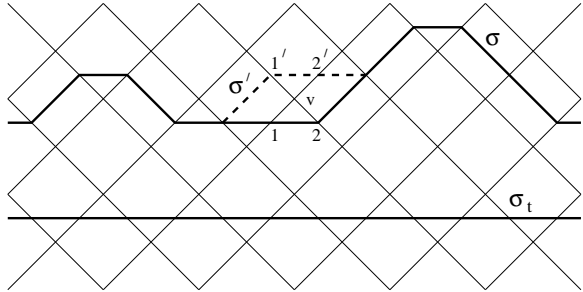}}
\caption{
The light-cone lattice. $\sigma_t$ is a constant time slice;
$\sigma$ is a general spacelike surface and $\sigma'$ one obtained
from it by an elementary motion across the vertex $v$. 
}
\label{lattice:fig}
\end{figure}

The local field variables, $\alpha$, live on the 
links. At link $l$ the variable $\alpha_l$ takes 
just two values, $0$ or $1$, and there is a (``qubit'') Hilbert 
space, ${\cal{H}}_l$, spanned by two states labelled by 
$\alpha_l=0$ and $\alpha_l=1$. 
At each vertex, $v$, the local evolution law is
given by a 4-dimensional unitary R-matrix, 
$U(v)$, that 
evolves from the 4-d Hilbert space that is the tensor
product of the Hilbert spaces on the two ingoing links
to the 4-d Hilbert space on the two outgoing links.  

A quantum state $|\Psi>$ on a spacelike surface,
$\sigma$, is an element of the Hilbert space,
${\cal{H}}_\sigma$, that is the 
tensor product of the Hilbert spaces on all the 
links cut by $\sigma$. $|\Psi>$ is  specified in the $\alpha$ basis as
a normalised complex function of the variables on the links cut by a 
spacelike surface, $\sigma$. Denoting this set of variables 
by $\alpha|_{\sigma}$, the wave function is written 
$\Psi(\alpha|_{\sigma})$. The unitary evolution of
the wavefunction to another spacelike surface $\sigma'$
is effected by applying all the R-matrices at the vertices 
between $\sigma$ and $\sigma'$, in an order respecting 
the causal order of the vertices.
In the simplest case, when only a single vertex is crossed 
(to the future of $\sigma$) the deformation of 
$\sigma$ to $\sigma'$ is called an ``elementary motion'' and 
an example is 
shown in figure 1. 

The R-matrices have been left unspecified to keep the
discussion as general as possible.
In a conventional field theory, they will be uniform over
the lattice. One particular choice and 
a suitable continuum limit leads, for example, to the 
massive Thirring model \cite{Destri:1987ze}. 

The standard interpretation of the theory is expressed
in terms of the results of measurements of any  
hermitian operator associated with any surface $\sigma$. 
The  state on $\sigma$ provides   the appropriate
probability distribution. This 
standard theory suffers from at least two problems. 
Firstly, it cannot be a theory of a closed system 
(the entire universe, say)
since it requires external, classical measuring agents. 
Secondly, serious
threats of superluminal 
signaling arise on trying to extend the 
interpretation to sequences of measurements 
tied to spacetime regions more general than hypersurfaces
(see {\it e.g.} \cite{Aharonov:1980aj, Sorkin:1993gg, Beckman:2001ck})
There are strong motivations  for trying to develop the field
theory into a realistic model in which  predictions would be  
observer independent and superluminal signaling does not 
occur.  
We will describe our attempt in the next section but 
first, for illumination and comparison, we give a  brief
description of another 
such model, the Samolsian dynamics.  

The Samols model is a realistic, stochastic model of 
the above lattice quantum field theory that agrees with the predictions
of the standard theory in situations where the latter makes
predictions.
The dynamics is defined inductively.
The initial conditions are that on some spacelike 
surface, $\sigma_0$, the wavefunction is $\Psi(\alpha|_{\sigma_0})$
and a configuration $\halpha|_{\sigma_0}$ is chosen at 
random according to the quantum mechanical probability 
distribution $|\Psi(\halpha|_{\sigma_0})|^2$.

Suppose we have a surface $\sigma_{k-1}$ with wavefunction and
some realised field configuration on it. 
One of the possible elementary motions occurs thus: 
at random one of the RL pairs is chosen from the sequence of links that 
defines the surface $\sigma_{k-1}$ and the surface is moved up across the 
the associated vertex so that the pair is replaced by LR. 
As in figure 1, let this motion be from links $1$ and $2$ to $1'$ and $2'$. 
The wavefunction is evolved forward to $\sigma_{k}$ by the 
R-matrix of the vertex.  Field values 
are realised randomly on the new links according to the 
conditional probability, 
$f_\Psi(\halpha|_{\sigma_{k-1} \rightarrow \sigma_{k}})$,
of realising values $(\halpha_{1'},\halpha_{2'})$ given all the 
realised values $\halpha_l$ up to then, where
\be
f_\Psi(\halpha|_{\sigma_{k-1} \rightarrow \sigma_{k}}) = 
\frac{|\Psi(\alpha|_{\sigma_{k}})|^2}{\sum_{\alpha_{1'}\alpha_{2'}}
|\Psi(\alpha|_{\sigma_{k}})|^2} 
\biggr\arrowvert_{\alpha |_{\sigma_{k}}=\hat\alpha |_{\sigma_{k}}}  
\ee
This rule guarantees that the marginal probability distribution on
$\alpha|_{\sigma_k}$ is equal to the quantum mechanical one.
It also means that there is no conditional dependence of 
$(\halpha_{1'},\halpha_{2'})$ on the realisations to the past 
of $\sigma_{k}$. 

For each sequence of hypersurfaces generated by possible
successive elementary motions, $\gamma= \{\sigma_0, \sigma_1, 
\sigma_2,\dots\}$,
this gives a 
probability distribution on the sample space of 
all field configurations (on and) to the
future of $\sigma_0$. To get the 
unconditional probability distribution we sum these over $\gamma$ with 
weights given by the stochastic rule for elementary motions stated above. 

The essential structure of  both the basic lattice 
quantum field theory and the Samolsian dynamics is
very simple and versatile. 
It requires only   
a discrete causal structure and a local unitary 
evolution, so the generalisation 
can easily made to the case of    
a quantum field theory on any 
locally finite partial order (a ``causal set'' \cite{Bombelli:1987aa})
where the field variables live on the links,  
as Samols describes.  

\section{GRW on the lattice}

One of the defining features of Samolsian dynamics is that 
it is operationally equivalent to standard quantum theory
in situations where the standard theory applies.
Someone who believes that standard quantum theory 
will never be found to give incorrect predictions
may consider this essential, but for those of us who keep  
a more open, scientific mind it is interesting 
to consider alternatives that give rise to 
predictions that  
differ from those of the standard theory. Spontaneous collapse
models  are such alternatives and so 
let us now construct a collapse 
version of the lattice field theory. 

In the original GRW  dynamics, the wave function is a function of
the position of the particle. When a collapse happens it is
centred on a particular, randomly chosen position and 
according to the Bell interpretation, that position at 
that time -- that {\it event} -- is then real.
We are considering a field theory here and the quantum state is a 
functional of the field configuration on a spacelike
surface. So, now, collapses will be centred on 
field  values and it will be one, randomly 
chosen field configuration on the lattice 
that will constitute reality in our
model, which proceeds inductively as follows. 

We start with a wave function $\Psi(\alpha|_{\sigma_0})$ 
on a spacelike
surface $\sigma_0$. 

1. Suppose we have $\Psi(\alpha|_{\sigma_{k-1}})$
on surface $\sigma_{k-1}$. At random, an elementary motion occurs and
the wavefunction $\Psi$ is evolved forward
by the unitary R-matrix
associated with the single vertex, $v_k$, crossed, to the new 
surface $\sigma_k$. There the resulting wavefunction  
is $\Psi(\alpha|_{\sigma_k})$.

2. A field value 
$\halpha_L$ is realised on the new L link. The value is chosen at
random from $\{0,1\}$ according to the GRW 
probability distribution $(N_L(\halpha_L))^2$ which will be 
defined shortly.

3. The wave function on the surface $\sigma_k$ 
suffers a ``hit'' and becomes

\be\label{collapse.eq} 
\Psi'(\alpha|_{\sigma_k}) = \frac{j_{\alpha_L \halpha_L} \Psi(\alpha|_{\sigma_k})}{N_L(\halpha_L)}
\ee
where the GRW ``jump factor'' is given by

\be\label{jump.eq}
j_{\alpha_L \halpha_L} = \frac{\delta_{\alpha_L \halpha_L} 
+ (1 - \delta_{\alpha_L \halpha_L}) X}{\sqrt{1+X^2}}
\ee

with $0\le X\le1$ and the normalisation given by

\be
(N_L(\halpha_L))^2 = \sum_{\alpha|_{\sigma_k}} j^2_{\alpha_L \halpha_L}
|\Psi(\alpha|_{\sigma_k})|^2
\ee
which is the probability distribution  in step 2. 

Just a word of explanation so that the notation is clear. The
link, $L$, is one of the links in $\sigma_k$ and in equation \eqref{collapse.eq}
$\alpha_L$ is therefore one of the field variables in the argument of the 
wavefunction.  For field configurations on 
$\sigma_k$ in which the variable $\alpha_L$ agrees with the value
$\halpha_L$ ({\it {i.e.}} the realised value) the amplitude 
for that field configuration is 
multiplied by the factor $1/(N_L \sqrt{1+X^2})$ and otherwise the amplitude 
is multiplied by $X/(N_L \sqrt{1+X^2})$.  
Thus the jump factor $j_L$ is chosen so that 
the amplitudes of field configurations that agree with 
the realised value $\halpha_L$ are enhanced over the 
amplitudes of field configurations that do not 
by the ratio $1/X$. For example, if $\halpha_L =1$ then 
the effect of the multiplication of the wave function by 
the jump factor is to act on the two dimensional 
Hilbert space for the link $L$ by the matrix
\be
\frac{1}{\sqrt{1+X^2}}
\begin{pmatrix}
1 & 0 \\
0 & X
\end{pmatrix}
\ee
where the first (last) row is labeled by the state 
with $\alpha_L =1$ ($\alpha_L =0$). 

4. A collapse occurs on the new right link, R, 
to a field value
$\halpha_R$. The value is chosen at
random from $\{0,1\}$ according to the
probability distribution $(N_R(\halpha_R))^2$.

5. The wave function on the surface $\sigma_k$
suffers a second ``hit'' and becomes
 
\be
\Psi''(\alpha|_{\sigma_k}) = \frac{j_{\alpha_R \halpha_R} \Psi'(\alpha|_{\sigma_k})}{N_R(\halpha_R)}
\ee
where
\be
j_{\alpha_R \halpha_R} = \frac{\delta_{\alpha_R \halpha_R}
+ (1 - \delta_{\alpha_R \halpha_R}) X}{\sqrt{1+X^2}}
\ee
and  
\be
(N_R(\halpha_R))^2 = \sum_{\alpha|_{\sigma_k}} j^2_{\alpha_R \halpha_R}
|\Psi'(\alpha|_{\sigma_k})|^2
\ee
 which is the probability distribution for step 4.  

6. Go to step 1 where now it is the wavefunction $\Psi''$ that is
evolved forward by the R-matrix of the next randomly chosen 
vertex.

It will be convenient in what follows to refer to 
the realisation of $\alpha$ values on the links R and L, outgoing 
from $v_k$, as a single 
{\it event at the vertex, $v_k$}.  
The values 
$\{\alpha_R, \alpha_L\}$ are summarised as $\alpha_{v_k}$.
The dynamics can then be re-expressed  as an elementary 
motion followed by a single
realisation of value  $\halpha_{v_k}$ with jump factor
\be\label{pairjump.eq}
j_{\alpha_{v_k} \halpha_{v_k}} \equiv  j_{\alpha_L \halpha_L} j_{\alpha_R
\halpha_R}
\ee
and probability distribution  
\be
(N(\halpha_{v_k}))^2 = \sum_{\alpha|_{\sigma_k}} j^2_{\alpha_v \halpha_v}
|\Psi(\alpha|_{\sigma_k})|^2 \ .
\ee

A given ``run'' of the dynamics will  
generate a random sequence, $\gamma = \{\sigma_1, \sigma_2, \dots \}$,
of surfaces to the 
future of the initial surface, $\sigma_0$, each related to the
previous one by an elementary motion. A sequence $\gamma$ is 
equivalent to a linear ordering of the vertices to the future of
the initial surface that is compatible with the causal 
order of the vertices (called a ``linear extension'' of the causal 
order or a ``natural 
labeling'' of the vertices). 

The probability distribution generated 
by this dynamics is, as in the Samolsian dynamics, a 
measure on the sample space, $\Omega$, 
of all possible field configurations to the 
future of $\sigma_0$. Strictly, what the dynamics 
gives is a probability measure on certain subsets of 
$\Omega$, the so-called ``cylinder sets.'' A cylinder set  
consists of all field configurations that agree with a given one on a 
``partial stem'' which is a finite set of vertices that contains its
own causal past (to the future of $\sigma_0$). Standard measure theory then 
guarantees that this extends to a measure on 
the $\sigma$-algebra generated by the cylinder sets, 
that is all sets formed by countable set operations on
the cylinder sets.

In the Samolsian dynamics, the probability distributions
that are conditioned on $\gamma$ are not equal. To 
obtain the full distribution, these are summed over all
$\gamma$ (all natural labelings of the 
vertices). Also, although the probability 
distribution on the possible events at a vertex ({\it i.e.} 
field values on a
single outgoing LR pair) is independent of the R-matrices 
spacelike to it, this is not true of the probability 
distribution on the events in a more general spacetime 
region.   

By contrast, and as anticipated by Samols, 
the enhanced locality property of our collapse model
means that the marginal probability distribution on 
any collection of events is independent of the 
R-matrices at vertices spacelike to the whole collection.
Moreover, the probability distributions conditioned on $\gamma$
are equal to each other. These claims are proved in the next 
section. This means that the order of evolution of the 
surfaces can be considered to be genuinely without 
physical meaning (in contrast to the 
Samolsian dynamics within which the sequence of 
hypersurfaces is without {\it operational} meaning 
because a local  observer cannot determine it). 
We can still, if we wish,
consider the  full probability 
distribution to be given by a sum over $\gamma$ of the 
distributions conditioned on $\gamma$ but the contribution 
from each is the same. This is  the analogue, in this setting, of 
general covariance: the independence of the action of an  
unphysical labeling.    

\section{The probability distribution is independent of the 
sequence of surfaces}
We show that the probability distribution is independent of 
the sequence of hypersurfaces.
Consider a sequence of surfaces, specified by 
a natural labeling of the vertices to the future of 
$\sigma_0$, $\{v_1,v_2,v_3,\dots\}$. Let the 
surface in the sequence just after the 
elementary motion across $v_k$ be denoted  $\sigma_k$ and the 
two outgoing links from $v_k$ be $L_k$ and $R_k$. 
The variable $\alpha_{v_k}$ stands for $\{\alpha_{L_k},\alpha_{R_k}\}$. 
Let the Hilbert space on $\sigma_k$ be denoted ${\cal{H}}_{\sigma_k}$ and
recall that it is the tensor product of $2N$ 2-d Hilbert spaces, one
for each link cut by $\sigma_k$.
The 
state 
on $\sigma_k$, just after the evolution
by the R-matrix at $v_{k-1}$ but before the hit, will be denoted 
by $|\Psi_k>$, and the state after the hit will be denoted by
$|\Psi'_k>$. $|\Psi_k>$ depends on the realised values $\{\halpha_{v_1},
\dots \halpha_{v_{k-1}}\}$ and $|\Psi'_k>$ 
depends on $\{\halpha_{v_1}, 
\dots \halpha_{v_k}\}$. 
They are related by
\be
|\Psi'_k> = \frac{J(\halpha_{v_k}) |\Psi_k> }{N_k(\halpha_{v_k})}
\ee
where $J(\halpha_{v_k})$ is the linear operator defined 
as follows.
$J(\halpha_{v_k})$ acts on the 
four-dimensional Hilbert space associated with 
the outgoing links from $v_k$ as the matrix:
\be 
J(\halpha_{v_k})_{\alpha_{L_k}\alpha_{R_k}\beta_{L_k}\beta_{R_k}}
= j_{\alpha_{L_k}\halpha_{L_k}} \delta_{\alpha_{L_k}\beta_{L_k}}
 j_{\alpha_{R_k}\halpha_{R_k}} \delta_{\alpha_{R_k}\beta_{R_k}} \quad
\text{(no sums)}\ .
\ee
with $j_{\alpha_{L_k}\halpha_{L_k}}$ given by equation \eqref{jump.eq}.
$J(\halpha_{v_k})$ acts as the identity on the other Hilbert spaces
in the tensor product $\cal{H}_{\sigma_k}$. 

Denote the R-matrix at $v_k$ by $U(v_k)$.

We claim that the probability that 
the field values $\{\halpha_{v_1},\dots \halpha_{v_n}\}$ 
are realised,
given the sequence of surfaces $\gamma =\{\sigma_1,\sigma_2\dots\}$, is 
\be\label{prob.eq}
P^\gamma(\halpha_{v_1},\dots \halpha_{v_n}) = 
|J(\halpha_{v_n})U(v_n)J(\halpha_{v_{n-1}}) 
U(v_{n-1})\dots J(\halpha_{v_1})U(v_1)|\Psi_0>|^2\ .
\ee

Proof of claim:

By induction.
The probability that $\alpha_{v_1} = \halpha_{v_1}$ is
\be
P^\gamma(\halpha_{v_1}) = 
|J(\halpha_{v_1})U(v_1)|\Psi_0>|^2\ .
\ee

Assume that equation \eqref{prob.eq} is true for
$n = k-1$. Then 
\be
P^\gamma(\halpha_{v_1},\dots \halpha_{v_k}) = 
P^\gamma(\halpha_{v_k}| \halpha_{v_1},\dots \halpha_{v_{k-1}})
P^\gamma(\halpha_{v_1},\dots \halpha_{v_{k-1}})
\ee
\be
 = {|J(\halpha_{v_k}) |\Psi_k>|^2}
{|J(\halpha_{v_{k-1}})
U(v_{k-1})\dots U(v_1)|\Psi_0>|^2} 
\ee
\be
 = \frac{|J(\halpha_{v_k}) U(v_k) J(\halpha_{v_{k-1}})|\Psi_{k-1}>|^2}
{|J(\halpha_{v_{k-1}})| \Psi_{k-1}>|^2}
|J(\halpha_{v_{k-1}})
U(v_{k-1})\dots U(v_1)|\Psi_0>|^2 \ .
\ee

We may replace $ |\Psi_{k-1}>$ in the numerator and
denominator of the fraction by 
$U(v_{k-1})J(\halpha_{v_{k-2}})\dots U(v_1)|\Psi_0>$ 
as the normalisation factors cancel out and the result follows.

From this we can see that the probability is unchanged by 
an exchange of the order of any two successive spacelike separated vertices
in $\gamma$, say $v_{l}$ and $v_{l+1}$, because
\be
[J(\halpha_{v_l})U(v_l), J(\halpha_{v_{l+1}})U(v_{l+1})] = 0
\ee
if
$v_l$ and  $v_{l+1}$ are spacelike. 
Any  order preserving list, $\gamma$, of 
finitely many vertices can be transformed into any 
other order preserving list, $\gamma'$, of the  
same vertices by a sequence 
of such exchanges. Thus, the probability distribution on 
$\{\halpha_{v_1},\dots \halpha_{v_{n}}\}$ is independent of
$\gamma$.

We use this to show that the model satisfies what we call
``external relativistic causality.'' By this we mean that 
external agents that exist in spacetime in addition to 
the field and that can affect the field only by 
changing the R-matrices locally  at their spacetime position
cannot use the field to send superluminal signals.
Suppose agent Alice is located in spacetime region $A$
and agent Bob in region $B$ such that all vertices
of $A$ are spacelike to all vertices  of $B$. Suppose 
Bob has some records of past events in the causal past 
of $B$. Can Alice signal to Bob by manipulating the 
R-matrices in $A$? This can only happen if the  
probability distribution on a set of events in $B$, conditional 
on some collection of events in $P(B)$ (the causal past of $B$),
 depends on an 
R-matrix in $A$.  This probability distribution can be calculated from
the joint distribution on all events in $B$ and $P(B)$, 
$ P(\halpha_B, \halpha_{P(B)}) $ where
$\halpha_B$ is shorthand for the $\alpha$ values in  
$B$ {\it etc.}. 
To calculate these probabilities we may use any natural labeling 
of the vertices to the future of $\sigma_0$. There exists a  natural
labeling that first labels the vertices in $P(B)$ and
then those in $B$. Since $A$ intersects neither $B$ nor $P(B)$,
we see that $ P(\halpha_B, \halpha_{P(B)})$ is independent
of the R-matrices in $A$.    

The further question arises of whether the model satisfies 
``internal relativistic causality'' where this would 
mean that if the field were the entire universe (with 
no external agents) no superluminal signaling 
could occur in that universe. This demands that a definition 
of ``superluminal signaling'' be made in this case. Such 
a definition does not exist, but preliminary versions are 
being worked on \cite{Dowker:2002}. 

We  may adopt a ``Heisenberg picture'' description and throw
the evolution onto the $J$'s by defining
\be
J_{v_k}(\halpha_{v_k}) = U^{-1}(v_1) \dots U^{-1}(v_k) J(\halpha_{v_k}) 
U(v_k) \dots U(v_1)\ .
\ee
Note that the R-matrices spacelike to $v_k$ can be commuted
through the expression and cancelled off 
with their inverses so that $J_{v_k}(\halpha_{v_k})$
only depends on the R-matrices in the causal past of $v_k$.
Then \eqref{prob.eq} becomes
\be\label{probtwo.eq}
P^\gamma(\halpha_{v_1},\dots \halpha_{v_n}) =
|J_{v_n}(\halpha_{v_n})
\dots J_{v_1}(\halpha_{v_1})|\Psi_0>|^2\ .
\ee

We see an immediate similarity with the 
form of the probability of a history in the consistent 
(or decoherent) histories approach to 
quantum mechanics due to Griffiths, Omn\`es, 
Gell-Mann and Hartle.
Differences with 
the consistent histories approach include the fact that 
the $J$ operators are, in general, not projectors (though they satisfy
$\sum_{\halpha_{v_k}} J^2(\halpha_{v_k}) = 1$ and are examples 
of what are known as ``Kraus operators'') 
and the histories are not ``consistent''. 

A  
translation of the non-relativistic GRW model into this ``historical''
framework was made by Kent 
\cite{Kent:1998bc}. Connections between 
consistent histories and collapse models and related theories of 
quantum open systems have also been made by
Diosi, Gisin, Halliwell and Percival and by Brun
\cite{Diosi:1995bp,Brun:1997mc,Brun:2001}.
It seems possible, following Kent, following
the consistent historians, to adopt the causally ordered list of 
$J$ operators, 
\be 
\{J_{v_1}(\halpha_{v_1}), J_{v_2}(\halpha_{v_{2}}),
\dots
J_{v_n}(\halpha_{v{n}})\}
\ee
 as the specification of the history of which 
equation \eqref{probtwo.eq} is the probability. That is, it appears that  
this is a possible alternative ontology for our model, different 
from the field configurations (the $J$'s depend on the
number $X$ for example, whereas the field configurations 
do not). Whether or not these are 
genuinely different ontologies, and what that would mean
if they nevertheless produce the same predictions, seems a subtle question, 
beyond the scope of the present paper. 

\section{Discussion}

The main value of our model is that it is very simple and 
straightforwardly illuminates many of the issues that arise in 
seeking realistic alternatives to standard quantum mechanics. 
The model has a high degree of
locality built into it and external agents cannot manipulate it to 
produce superluminal signals. However, there 
{\it are} ``non-local correlations'':
the probability distribution on events at a vertex
will generally depend on the 
events realised at vertices spacelike to it. 
In Rideout and Sorkin's terminology, 
the model does not satisfy ``Bell causality'' 
\cite{Rideout:1999ub}
and so it has the potential of reproducing the 
Bell-inequality-violating correlations that most  
likely occur in nature. 

The simplicity of \eqref{probtwo.eq} and its 
connection to the consistent histories formalism suggest
numerous potential variations: 
different choices of the jump or Kraus operators, 
branch dependence (where the choice of $J$'s at a vertex can depend on 
events in its causal past), using a mixed state instead 
of a pure state, and adding a final state.  

With a given choice of R-matrices, the 
continuum limit may be examined. This would be 
done by a procedure of coarse graining and 
renormalisation of the $X$
parameter, studying the limit of the 
probability distribution to see if it tends to 
something well-defined.   With some 
choices, it is possible that the
continuum limit could be one of the models studied by
Adler and Brun \cite{Adler:2001} as these share
one of the main features of our lattice model, namely
that it is constructed using a locally, randomly 
evolving surface. The Adler-Brun models display 
an unphysical large energy 
production and we can ask whether our lattice model can 
be  adjusted so as to avoid this. 
For example,  one variation  of the model 
would be to let the realisation of a field value 
on each pair of links have only a probability of 
occurring which would have the effect of making
the field configuration sparser. The probability of 
realisation at a vertex could be fixed or could vary according
to events in the causal past. 

Because of its extreme simplicity, the model 
seems well suited to exploring the issue
of ``internal'' relativistic 
causality, as mentioned above. That is, we can try to 
formulate a definition of internal relativistic causality 
for the model -- some  condition on the probability 
distribution on field configurations -- and then see if it is 
satisfied. It may be, however, that this definition will also 
depend on the relationship of the field configurations to the 
macroscopic world of experiments and experimenters and will
be hard to glean without this having been determined.

This is connected to the major question: what, if anything, do these
models describe physically?  Do the realised field
configurations exhibit interesting behaviour or are they 
just  too noisy to see any structure 
%
%
(note that when the parameter
$X$ is chosen to be 1, the evolution of the state is the standard
quantum mechanical unitary evolution and the unrelated probability 
distribution on the field values is that on a set of independent 
variables which are 0 or 1 with probability  $\frac{1}{2}$). 
Do superpositions of macroscopically distinct configurations --
whatever that means in this context -- collapse onto one
or other of those configurations, as we would want? 
Simulations are being done to investigate these
questions \cite{DowkerHerbauts:2003}.

These simulations may also throw light on the issue, 
raised by Kent \cite{Kent:1989nk}, of the status of the quantum 
state in collapse models with the Bell interpretation. 
We wish to say that the  real variables
are the field values alone.
For this to be a satisfying interpretation, it seems necessary, 
as Kent stresses, 
that the quantum state on any spacelike surface be equivalent 
to the historical record of events to the past of that 
surface. If it were not so then the information about the
quantum state  would be 
needed, over and above the information about past and present events, 
in order to be able to make predictions about the future. 
Denying the reality of the quantum state would then be an awkward thing  
to have to do. 
The simulations in progress will test the hypothesis that 
in the limit of late times, the probability distribution on 
future events depends only on past and present events and not on the 
quantum state on the initial hypersurface. 
We could then interpret the quantum state on a given hypersurface
as a useful fiction, an executive summary of the past that allows
future prediction. If this turns out not to be the case, however, 
one could still consider the model using the interpretation in   
which the state is taken to be real.

The null lattice used here is special to 1+1 dimensions
and this presents a difficulty in generalising our model
to higher dimensions. However, from the point of view of
the causal set approach to quantum gravity, spacetime is 
a continuum approximation to a discrete underlying substructure
(reality) and this substructure is a causal set, which, as mentioned
previously, is a locally finite partial order. In the 
case of a continuum approximation that is $d+1$ dimensional
Minkowski spacetime (perhaps identified on a $d$-torus) 
then the underlying reality is a causal set that can be 
produced by sprinkling points into the spacetime 
by a unit density Poisson process ({\it i.e.} 
the mean number of points in any 
region is equal to the volume of the region in Planck units) 
and endowing them with the partial order that they inherit 
from the spacetime causal order. The rules of 
a collapse model on a background causal set would 
involve putting field variables and associated Hilbert spaces
on the links and unitary R-matrices 
at the vertices -- as described by Samols \cite{Samols:1995zz}
-- and supplementing this by placing jump (Kraus) operators 
on the links also.  
This seems feasible.
If the observer independent successor theory to 
standard quantum theory 
turns out to be something along these lines, involving the 
discrete causal structure that underpins the 
spacetime continuum approximation, it might then be 
said that the 
resolution of the measurement problem does indeed 
involve gravity as many workers have suggested.

\section{Acknowledgments}

We are grateful to Adrian Kent,
Wayne Myrvold, Trevor Samols and Antony Valentini for 
interesting and helpful discussions. This work was supported
by the Department of Physics at Queen Mary, University of 
of London where much of it was completed. 


\end{document}